# Field-induced rocking curve effects in attosecond electron diffraction

Y. Morimoto[1,2,*] and P. Baum[1,*]

[1] *Universität Konstanz, Fachbereich Physik, 78464 Konstanz, Germany*
[2] *RIKEN Cluster for Pioneering Research and RIKEN Center for Advanced Photonics, 351-0198, Wako, Saitama, Japan*

*peter.baum@uni-konstanz.de
* yuya.morimoto@riken.jp

**Recent advances in electron microscopy trigger the question whether attosecond electron diffraction can resolve atomic-scale electron dynamics in crystalline materials in space and time. Here we explore the physics of the relevant electron-lattice scattering process in the time domain. We drive a single-crystalline silicon membrane with the optical cycles of near-infrared laser light and use attosecond electron pulses to produce electron diffraction patterns as a function of delay. For all Bragg spots, we observe time-dependent intensity changes and position shifts that are correlated with a time delay of 0.5-1.2 fs. For single-cycle excitation pulses with strong peak intensity, the correlations become nonlinear. Origin of these effects are local and integrated beam deflections by the optical electric and magnetic fields at the crystal membrane that modify the diffraction intensities in addition to the atomic structure factor dynamics by time-dependent rocking-curve effects. However, the measured time delays and symmetries allow to disentangle both effects. Future attosecond electron diffraction and microscopy experiments need to be based on these results.**

Femtosecond and attosecond pump-probe experiments play a pivotal role in understanding the properties of complex materials and their ultrafast reaction paths. In particular, pulsed electron beams [1–3] and ultrafast x-ray sources [4,5] have wavelengths that are shorter than atomic distances and can therefore resolve structural dynamics in space and time. However, the primary response of a material to light is given by the motion of electrons in the electromagnetic excitation wave on time scales as short as attoseconds. In order to see such dynamics, ultrafast electron microscopy has recently been advanced from the femtosecond into the attosecond domain [6,7], based on pioneering concepts for laser-electron control [6–18]. However, the spatial resolution is not at atomic dimensions yet, and additional efforts are therefore appropriate.

A potential solution for merging sub-atomic resolution in space with attosecond information in time is attosecond electron diffraction. Figure 1(a) shows the basics of such an experiment. A crystalline material (green) is excited by the electric field cycles of laser light (red), pushing and pulling electron densities between the atoms [19] and along the chemical bonds [20]. During this motion, synchronized attosecond electron pulses [6,10–15] (blue) are applied to produce an electron diffraction pattern as a function of time delay. In theory [19], the diffraction intensities should then reveal the electronic motion via a time-frozen structure factor [19,21–23].

However, all simulations and preliminary experiments [6] so far ignore the unavoidable presence of the optical excitation fields in front and behind the material which modulate the space-time profile of electron beams [19,21,22]. While the Kroll-Watson formula [24] guarantees for isolated atoms a safe extraction of the form factor from energy-integrated diffraction intensities, there is no such theorem on the diffraction by crystals with periodic atoms in the Bragg diffraction regime. It therefore remains to be resolved how electron diffraction from condensed matter works on attosecond time scales and how time-dependent Bragg spot intensity changes relate to the attosecond dynamics of the investigated material and its internal optical fields.

Figure 1(a) shows the basics of our experiment. The material under investigation (green) is a 60-nm thick single-crystalline membrane of silicon (Norcada) under excitation of the optical field cycles of femtosecond laser light at $\lambda = 1030$ nm wavelength (red). Attosecond electron pulses (blue) diffracted from the crystal lattice into Bragg spots on a detector screen (grey). In more detail, we first create femtosecond electron pulses by two-photon photoemission [25] and accelerate them to a central energy of 70 keV. A train of attosecond electron pulses (blue) is produced by energy modulation at a 50-nm-thick silicon nitride membrane [6]. The pulse duration is 0.8 fs and the electron velocity is $v_e \approx 0.48c$. The number of electrons per pulse train is less than one [26] in order to avoid space charge effects, and the divergence of the electron beam is about 0.05 mrad. Each attosecond micro-bunch (blue) is separated in time by an optical cycle period (3.4 fs). The attosecond electron beam hits the silicon membrane at an angle of ~35° under a surface normal in $[-\frac{1}{\sqrt{2}}, \sqrt{2}, \frac{1}{\sqrt{2}}]$ direction. A single-electron-sensitive camera system [27] is applied at 1.3 m distance for electron detection.

Figure 1(b) shows the resulting Bragg diffraction pattern that we obtain with our attosecond electron pulses in case of no laser excitation. We see Bragg spots at a large variety of Miller indices. We then excite the silicon crystal by the pump laser field under *p*-polarization and 145° incidence angle with respect to the electron beam. The optical pulse duration is 1.7 ps and the electric peak field strength is $F = 0.2$ V/nm. Figure 1(c) reveals a periodic deflection of the Bragg spots and the direct beam on attosecond time scales. Figure 1(d) shows the evaluated angle changes as a function of time. Analogously, we also evaluate the Bragg spot intensities as a function of time. Figure 1(e) shows a typical result at the example of the $1\bar{1}3$ spot. We find a substantial intensity modulation of almost one percent that is reproducible over multiple pump-probe scans. The oscillations have the same period as the deflection (3.4 fs) but appear with a substantial time delay (red marks).

Figure 1(f) shows a correlation plot between the measured intensity changes $\Delta I_{1\bar{1}3}(t)$ and the measured sideways deflections $\alpha_{\text{fin}}(t)$, accumulated over all attosecond time delays. We see an elliptical pattern with a tilt (blue). Positively deflected Bragg spots have generally less intensity than negatively deflected spots, but the correlation is not direct: unchanged intensities, for example, occur at two distinct deflections and time delays. A least-square fitting (blue) with two phase-shifted sinusoidal oscillation curves [compare Fig. 1(d)-(e)] reveals a time delay of $-1.17 \pm 0.08$ fs or a phase delay of $-2.14 \pm 0.07$ rad between the beam deflection $\alpha_{\text{fin}}(t)$ and the Bragg intensity $\Delta I_{1\bar{1}3}(t)$. These values are substantial and far away from any multiple of $\pi$ or $\pi/2$; in other words, the measured ellipse is neither a line nor a circle. The other seven Bragg spots show a similar behavior (see below).

These results show directly that Bragg diffraction with attosecond electron pulses from a laser-driven material is subject to substantial and complex beam dynamics that is linked to the optical cycles of the excitation light. In order to understand the measured attosecond beam deflections, we look at the electron trajectories close to the specimen; see Fig. 2(a). While the attosecond electron pulses (blue) pass through the optical focus (red) and the silicon crystal (green), they accumulate from the time-integrated electromagnetic fields of the excitation laser a final sideways deflection (blue) that oscillates as a function of the arrival time [6,8], because the material breaks the symmetry of a free-space interaction and enables electron-photon momentum exchange [8,28–31]. Consequently, the direct electron beam and all far-field Bragg diffractions rapidly oscillate on the screen as a function of their initial arrival time [6].

In order to understand the measured attosecond intensity changes, we first estimate what effects can be expected from atomic-scale electron dynamics [19]. Density-functional theory with a static-field approximation [6] reveals that the expected intensity changes for our material and laser intensity are merely 0.01% and should appear at a frequency of twice that of the excitation-field [19], because left-driven atomic charges and right-driven atomic charges produce the same Bragg-spot changes due the symmetry of Friedel pairs. However, the measured Bragg spot changes appear at the fundamental laser frequency. Our observed intensity oscillations can therefore not be explained by changes of the scattering form factor and field-driven electronic motions on atomic dimensions.

Instead, the scattering process itself is modified by the presence of the excitation light. As illustrated in Fig. 2(a), the optical laser fields (red) not only excite the specimen but also induce an unavoidable quiver motion of the incoming attosecond electron pulses (blue) before, within and behind the diffracting material (green). Although the physical distances that the electrons travel away from the optical axis are very small ($eF/m\omega^2 < 1$ nm, where $m$ is electron's mass, $e$ is unit charge and $\omega = 2\pi c/\lambda$), the electrons pass through the specimen at special instantaneous angles that are determined by the specimen and the optical geometry. Electric and magnetic fields are both relevant for this dynamics. In our experiment, the attosecond electron pulses receive oscillating accelerations and decelerations in the $x$-$z$-plane and enter the crystal at an angle $\alpha_{\text{sample}}(t)$ that is determined by the integrated laser fields of the left half of the geometry. After interacting with the specimen, the electrons see continuing oscillations in the remaining laser fields and acquire their final far-field deflection [8,28,31,32].

Electron diffraction at a non-optimized angle of incidence causes Bragg spot attenuation or amplification due to rocking curve effects. Figure 2(b) depicts the reciprocal crystal lattice (black) together with the electron beam's Ewald sphere (blue). The reciprocal spots are broadened by temperature and additionally in $z$ direction by finite crystal thickness. Provided that the time it takes to diffract into Bragg spots ($d/v_e \approx 0.4$ fs) is shorter than half an optical cycle period (1.7 fs), the local angle $\alpha_{\text{sample}}(t)$ is approximately well-defined. An oscillating $\alpha_{\text{sample}}(t)$ on attosecond dimensions (dotted blue lines) therefore creates a rapidly quivering Ewald sphere (dashed blue circles) that rotates periodically around the origin O at the frequency of the laser light. This ultrafast rotation increases or decreases the overlap with the reciprocal lattice points and modulates the measured diffraction intensities. Importantly, the oscillation period is the same as the laser period, and opposite Bragg spots (Friedel pairs) obtain approximately opposite effects.

Figure 2(c) depicts several measured rocking curves in our experiment, that is, measured electron diffraction intensities under systematic variation of the angle of incidence ($\alpha_{\text{rock}}$). Attosecond electron pulses or a continuous beam yield almost identical results. We see for the $1\bar{1}3$ Bragg spot an almost Gaussian curve while the $\bar{1}\bar{1}1$ and 202 spots, for example, have dips in the middle due to multiple scattering effects [33]. The black dotted lines are the static geometric angles of the attosecond experiment, and the red arrows indicate the rapid oscillation of $\alpha_{\text{sample}}(t)$ due to the optical cycles of the pump laser field. The slope and curvature of the rocking curve of the specimen at the chosen diffraction geometry therefore translates the attosecond quivering of the electron beam at the instance of diffraction into attosecond modulations of measured Bragg spot intensities. Afterwards, the entire diffraction pattern continues to oscillate in the remaining laser fields until an angularly oscillating far-field pattern is produced.

To relate $\alpha_{\text{sample}}(t)$ and $\alpha_{\text{fin}}(t)$ by theory and confirm these explanations, we consider the incident electrons and the laser fields as a point particles and plane waves, respectively [8,32]. Optical thin-film interferences are taken into full account. Figure 2(c) shows the simulated $\alpha_{\text{sample}}(t)$ and $\alpha_{\text{fin}}(t)$ and their correlations at three essential points of the specimen, that is, at the front surface, in the middle and at the rear surface. All ellipses show substantial areas and therefore time delays. However, they are not all equal, because the time of electron diffraction ($d/v_e = 0.4$ fs) is not entirely irrelevant with respect to the excitation laser's cycle period (3.4 fs). In the experiment, we measure an average of all three dynamics and therefore consider here the middle one (green) for further analysis. The theoretical delay between $\alpha_{\text{sample}}(t)$ and $\alpha_{\text{fin}}(t)$ is 0.48 fs or 0.88 rad. According to Fig. 2(c), the rocking curve in this experiment has a negative slope, that it, a positive $\alpha_{\text{sample}}$ decreases the diffraction intensity. The delay between $\Delta I(t)$ and $\alpha_{\text{fin}}(t)$ is therefore 0.88 rad $- \pi = -2.3$ rad. These values match well to the experiment (−2.1 rad or −1.2 fs).

To obtain a more systematic picture, we repeat our experiments for the other seven more diffraction spots. For each Bragg spot, the crystal is aligned slightly away from the optimum Bragg condition by 0.07±0.05 mrad which is accuracy of our goniometer mechanics. Through the same analysis as above, we obtain for each Bragg spot the time delays and amplitudes of the deflection and intensity modulations. Figure 2(f) shows the extracted time delays. These delays accumulate around −1.2 fs and +0.5 fs (green lines). The negative delays correspond to the theoretical prediction of −2.26 rad (lower green line), indicating Bragg spot misalignments on negative slopes of the rocking curve (for example $1\bar{1}3$ in Fig. 2c), while the positive delays correspond to Bragg spots with positive rocking curve dynamics (for example 202 in Fig. 2c).

Figure 2(f) shows the measured intensity oscillation amplitudes for each Bragg spot as a function of the slope of the corresponding rocking curve, post-characterized after each attosecond measurement. The results are approximately on a straight line (green) because $|\alpha_{\text{sample}}|$ is small in this experiment (<0.1 mrad) and an oscillating $\alpha_{\text{sample}}(t)$ of constant magnitude therefore translates into intensity variations only by the strength of the local slopes of the rocking curves. We see that intensity amplitudes are typically smaller at lower-order Bragg spots such as 202 and $11\bar{1}$, because their rocking curves are rather broad; see Fig. 2(b). On the other hand, for Bragg spots such as $1\bar{1}3$ and similar diffraction orders, small accuracy-limited derivations of the specimen alignment ($\alpha_{\text{rock}}$) typically give a larger slope, and we observe stronger intensity modulations.

Before reporting more experiments, we make some intermediate conclusions: First, the observed attosecond modulations of Bragg spot intensities via $\alpha_{\text{sample}}(t)$ are not a specific result of our experimental geometry but general for any experiments in which a laser excitation has electric field components perpendicular to the electron beam direction in order to drive atomic-scale dynamics in perpendicular direction to the beam [19]. Whenever there are sideways fields there are also attosecond rocking curve effects. Second, the modulation amplitude of $\alpha_{\text{sample}}(t)$ scales proportionally to $eF/\omega p_e$ [8,34], that is, the diffraction intensity modulation is stronger with higher field amplitude, weaker for shorter excitation wavelengths and stronger for slower electrons. Also, the measured rocking curve effects are stronger for thicker crystals because rocking curves usually become sharper in proportion to the longitudinal number of unit cells in the specimen. These considerations will be helpful for designing an appropriate experiment. Third, the reported rocking curve effects are stronger for higher-order Bragg diffractions because the shift of the Ewald sphere, $\Delta q_z(q_x, t) \approx \alpha_{\text{sample}}(t) q_x$ in Fig. 2(b), is larger there. This result is in contrast to the expected modulations from the atomic scattering form factors due to electronic motion which is stronger at lower Miller indices due to the dominant contribution of outer electrons afar from the cores [19]. A combined measurement of multiple Bragg diffraction orders can therefore discern these two effects.

In a next experiment, we use much stronger single-cycle excitation pulses and explore the onset of nonlinearities in attosecond electron diffraction in strong laser fields. Figure 3(a) depicts the new experimental geometry. The laser pump pulses (red) are now obtained from a mid-infrared optical parametric amplifier [35] which provides ~20 times shorter pulses (36 fs) and ~7 times longer optical cycles (center wavelength 6.9 µm) than before. The carrier-envelope phase is not locked. The electron pulses at a duration of ~800 fs [36] are now longer than the laser pulses and we therefore perform here a time-integrated experiment where deflection data is averaged over all time delays, in a similar way as in free-electron tomography [37]. Nevertheless, each measured beam deflection still can be assigned to a measured diffraction intensity, and it is therefore possible to extract correlations and rocking curve dynamics from such an experiment. Figure 3(b) shows the measured time-integrated far-field streaking of the $\bar{1}1\bar{3}$ Bragg spot (upper panel) in comparison to the dynamics of the direct beam (lower panel). With pump laser ($F$ = 0.5 V/nm), both beams obtain a substantial broadening in $x$ direction (right panels). Strong signals at $\alpha_{\text{fin}} \approx 0$ are electrons which did not interact with any substantial part of the laser waveform due to the temporal mismatch [37]. We therefore focus our analysis on signals at final angles of $|\alpha_{\text{fin}}| > 0.1$ mrad. Figure 3(c) shows the measured rocking curve of the $\bar{1}1\bar{3}$ Bragg spot and the two dotted lines indicate two intentional misalignments that we apply in our experiments. For a rather low misalignment of 0.9 mrad, Fig. 3(d) shows the simulated correlation diagram (green line). We see a strongly distorted ellipse with one side even bending backward to the inside. The laser pulses are now so strong that the oscillations of $\alpha_{\text{sample}}(t)$ shown in Fig. 2(e) are almost as large as the rocking curve width and therefore drive for some delays the instantaneous electron diffraction over the top into regions of negative slope. The right panel of Fig. 3(d) shows the measured diffraction intensities $\Delta I$ as a function of $\alpha_{\text{fin}}$. As shown by the blue dotted line, there are two timings per laser cycle (blue dots) that result in the same $\alpha_{\text{fin}}$. The average of these two instantaneous intensities gives the time-integrated diffraction intensity at $\alpha_{\text{fin}}$ that is seen in the experiment; remember the absence of time resolution [37]. We see a reasonable agreement, demonstrating that attosecond rocking-curve effects are substantial in strong-field-driven electron diffraction experiments even if no compressed electron pulses are applied.

Figure 3(e) shows the results when we align the silicon crystal angle intentionally into a local minimum of the rocking curve; see right dotted line in Fig. 3(c). The left panel again shows the simulation results (green curve), revealing approximately a Lissajous figure with the laser frequency in $\alpha_{\text{fin}}(t)$ direction and the sum of its second and third harmonics in $\Delta I(t)$ direction.

Origin of this behavior is the quadratic (symmetric) and cubic (anti-symmetric) responses of the rocking curve at the chosen position with respect to the field-linear deflection angles $\alpha_{\text{sample}}(t)$. The right panel shows the measured correlation data (black circles) in comparison to $\alpha_{\text{fin}}$ (green trace). Again, the simulation matches to the experimental results, demonstrating that nonlinearity of the rocking curve is translated into nonlinear intensity dynamics in strong-field electron diffraction experiments.

In combination, these results show that the scattering process in electron diffraction from crystalline materials under excitation by laser light is affected by the unavoidable instantaneous electric and magnetic fields of the optical cycles within the material and around. Electron-lattice scattering is subject to substantial and nonlinear rocking curve effects that modulate the electron beam deflection and all Bragg spot intensities with substantial time delays. The origin of these delays are differences between the instantaneous and the time-integrated optical fields of the excitation laser along the electron trajectory before and after Bragg diffraction. Fortunately, the here presented results allow to disentangle the attosecond dynamics of the electron-lattice scattering process from the atomic structure-factor effects by recording multiple Bragg spots or entire rocking curves as a function of the attosecond time delay. Attosecond electron diffraction therefore remains a feasible and useful next step for advancing attosecond electron imaging to the atomic resolution regime.

**Acknowledgements:** This research was supported by the German Research Foundation (DFG) via SFB 1432, the Vector foundation, the Dr. K. H. Eberle Foundation, JST ACT-X JPMJAX21AO, JST FOREST, MEXT/JSPS KAKENHI JP21K21344, the Kazato Research Foundation and the Research Foundation for Opto-Science and Technology.

**Data availability:** The data supporting the findings of this study are available from the corresponding author upon reasonable request.

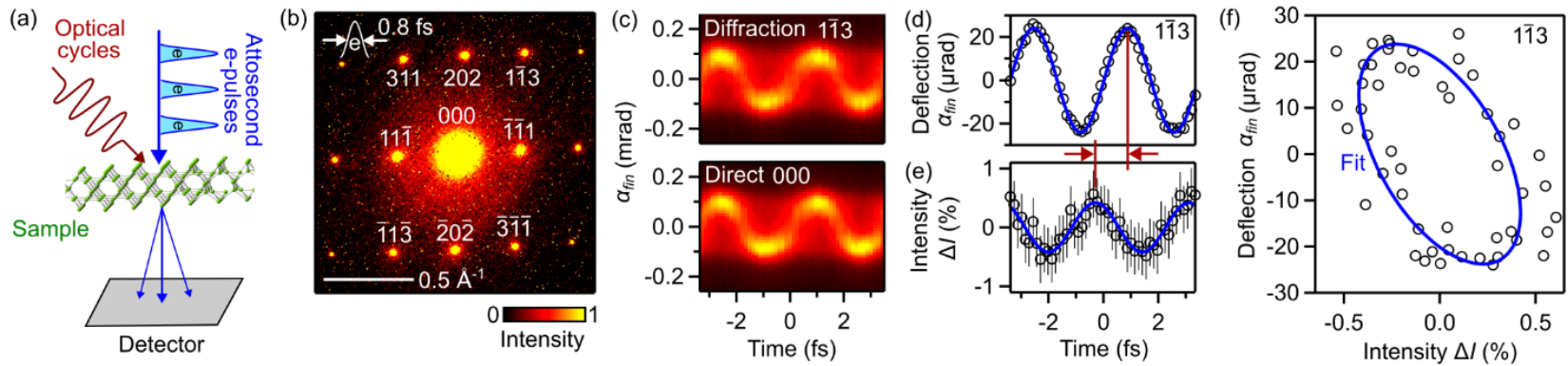

Fig. 1: Attosecond electron diffraction and measured time delays. (a) Concept of attosecond electron diffraction. A crystalline material (green) is excited by the optical cycles of laser light (red) and then investigated with attosecond electron pulses (blue). (b) Attosecond diffraction pattern of a Si crystal without laser excitation. The electron pulse length is 0.8 fs (white). (c) Raw data of the observed Bragg spot dynamics as a function of delay time. (d) Analysis of the time-resolved center-of-mass motion of the $1\bar{1}3$ Bragg spot. Analysis of the time-resolved diffraction intensity. The red lines and arrows indicate a time delay. (f) Correlation between sideways streaking (vertical axis) with the diffraction intensity (horizontal axis). Blue line is a double-sinusoidal fit.

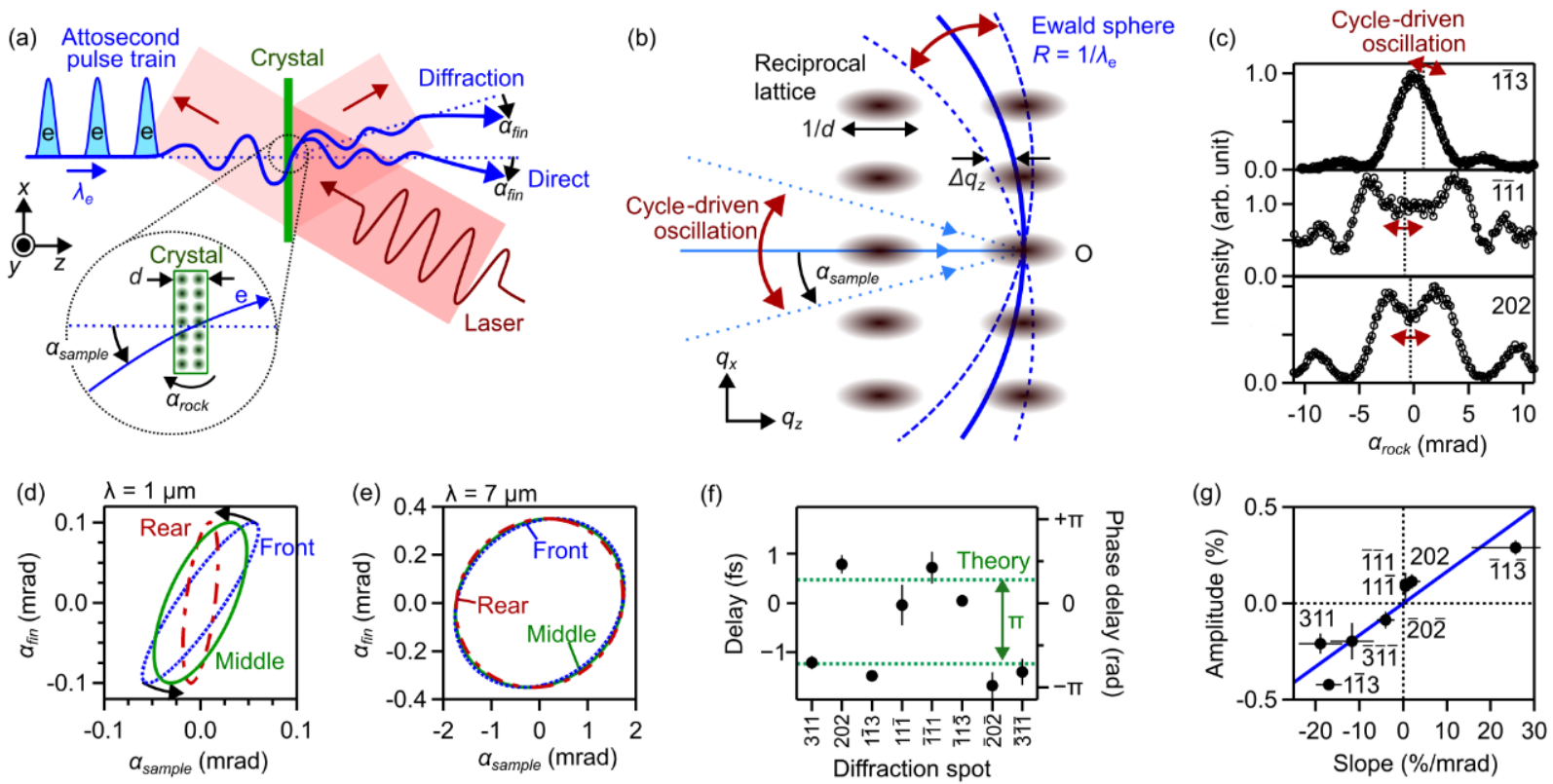

Fig. 2: Attosecond rocking curve effect. (a) Details of the experiment. $\alpha_{\text{sample}}$, instantaneous beam deflection; $\alpha_{\text{fin}}$, final beam deflection; $\alpha_{\text{rock}}$, static material angle; $\lambda_e$, electron de Broglie wavelength; $d$, crystal thickness. (b) Reciprocal-space dynamics. The cycle-driven rotation of the Ewald sphere (blue) modulates the overlap with the reciprocal lattice (black). $R$, radius of the Ewald sphere; O, origin; $\Delta q_z$, shift of the Ewald sphere along $q_z$. (c) Measured rocking curves for three diffraction spots. (d) Simulated phase delay of $\alpha_{\text{sample}}(t)$ with respect to $\alpha_{\text{fin}}(t)$ for 1-µm laser light. (e) Simulated phase delay of $\alpha_{\text{sample}}(t)$ with respect to $\alpha_{\text{fin}}(t)$ for 7-µm laser light. (f) Observed time delays for eight different Bragg diffraction spots. (g) Measured intensity modulation amplitudes and their correlation with the measured rocking-curve slopes. Blue line, linear fit.

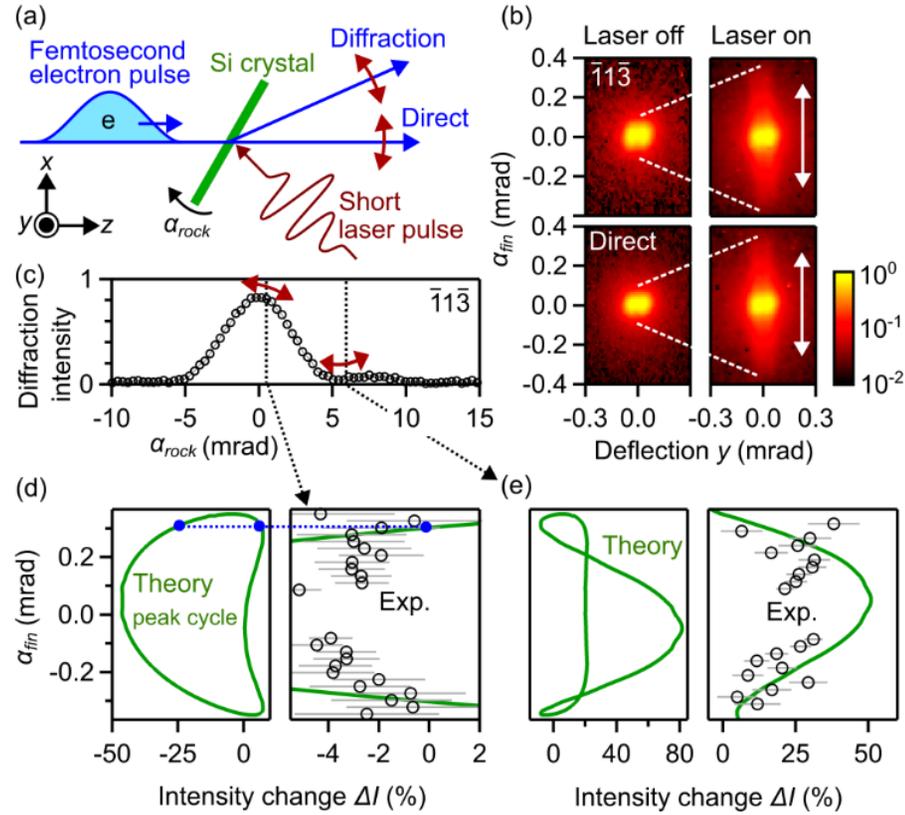

Fig. 3: Nonlinear modulation of diffraction intensities with infrared single-cycle light. (a) Schematic of the experiment. (b) Measured time-integrated sideways streaking (white arrows) of the $1\bar{1}3$ spot (upper panels) and the direct beam (lower panels). (c) Measured rocking curve of the $1\bar{1}3$ spot. The black dotted lines and arrows denote two conditions of the experiment with their corresponding attosecond dynamics. (d) Simulated correlation of diffraction intensity and final streaking angle (green) from the excitation peak at $\alpha_{rock} = 0.9$ mrad and comparison with the experimental results (black circles). Blue dots show two timings leading to the same $\alpha_{fin}$. Analysis is restricted to the strongest optical cycle of our excitation field because secondary cycles have only ≤60% field strength and $|\alpha_{fin}| \leq 0.2$ mrad. (e) Simulated correlation of diffraction intensity and final streaking angle (green) from the excitation peak at $\alpha_{rock} = 6$ mrad and comparison with the experimental results (black circles). Weaker optical cycles are neglected here.